\documentclass[10pt]{article}
\usepackage[dvips]{graphicx}

\usepackage[dvips]{graphicx}
\usepackage{amssymb,amsfonts,amsmath}
\usepackage[utf8]{inputenc}

\usepackage{overpic}
\DeclareSymbolFont{extraup}{U}{zavm}{m}{n}
\DeclareMathSymbol{\vardiamond}{\mathalpha}{extraup}{87}
\usepackage{times}

\usepackage{setspace}
\usepackage{caption}
\title{Food web assembly rules.}

\author{{\bf Jan O. Haerter, Namiko Mitarai, Kim Sneppen} \\
\vspace{.3cm}\\
Center for Models of Life,
Niels Bohr Institute, University of Copenhagen\\
Blegdamsvej 17, DK-2100, Copenhagen, Denmark
}

\date{}

\begin{document}

\baselineskip24pt

\maketitle

\noindent
{\bf corresponding author}: Jan O. Haerter, Center for Models of Life, Niels Bohr Institute, University of Copenhagen, Blegdamsvej 17, DK-2100, Copenhagen, Denmark, tel: +45 353 25352, fax: +45 353 25425, email: haerter@nbi.dk.\\
\noindent
{\bf Statement of authorship}: All authors developed the theory, performed the data analysis and wrote the manuscript and supplementary information.\\
\noindent
{\bf Word count} (excluding abstract, acknowledgments, tables, figure legends, and references): approx. 4500 words.\\
\noindent
{\bf Number of figures}: 6 figures.\\
\noindent
{\bf Conflict of interest}: The authors have no conflicting interests.\\
\noindent
{\bf Keywords}: Food web, competitive exclusion, apparent competition, biodiversity, parasite-host, concomitant link, food web stability, sustainability, trophic levels, food chain length.

\pagebreak

\noindent
\section*{ABSTRACT}\label{sec:abstract}
{\bf
Fueled by ongoing rapid decline of biodiversity \cite{ricciardi1999extinction}, ecology is in the midst of a lively debate on the effect of species loss or introduction on food web stability \cite{mccann2000diversity,montoya2006ecological}.
In food webs, many interacting species coexist despite the restrictions imposed by the competitive exclusion principle \cite{gause1936struggle,hardin1960competitive} and apparent competition \cite{holt1977predation}. 
For the generalized Lotka-Volterra equations \cite{campbell1960}, sustainable coexistence necessitates nonzero determinant of the interaction matrix. 
Here we show that this requirement is equivalent to demanding that each species be part of a non-overlapping pairing, which substantially constrains the food web structure.
We demonstrate that a stable food web can always be obtained if a non-overlapping pairing exists.
If it does not, the matrix rank can be used to quantify the lack of niches, corresponding to unpaired species.
For the species richness at each trophic level, we derive the food web assembly rules, which specify sustainable combinations.
In neighboring levels, these rules allow the higher level to avert competitive exclusion at the lower, thereby incorporating apparent competition.
In agreement with data \cite{Martinez1991,dunne2004network}, the assembly rules predict high species numbers at intermediate levels and thinning at the top and bottom. 
Using comprehensive food web data \cite{lafferty2006parasites,dunne2013parasites}, we demonstrate how omnivores or parasites with hosts at multiple trophic levels can loosen the constraints and help obtain coexistence in food webs.
Hence, omnivory may be the glue that keeps communities intact even under extinction or ecological release of species.
}
\pagebreak


\noindent
\section*{INTRODUCTION}\label{sec:introduction}
In food webs, 
complexity arises from combining a large number of species (the nodes) and a large number of relations between these species (the links).
Addressing the latter, recent attention was devoted to the structure of links using  e.g. the random, cascade and niche models \cite{cohen1985stochastic,williams2000simple}, stirring a prolific debate on the role of the link distribution regarding food web stability \cite{mccann2000diversity,montoya2006ecological,ives2007stability,allesina2012stability}.
We take a complementary approach: Using standard consumer-resource equations, we demonstrate fundamental constraints on node diversity in a food web, termed food web assembly rules.

For consumer-resource relationships, the competitive exclusion principle states that 
when two consumers compete for the exact same resource within an environment, one consumer will eventually outcompete and displace the other \cite{gause1936struggle,hardin1960competitive}.
It is known that the number of coexisting species cannot exceed the number of resources these species compete for \cite{macarthur1964competition}.
Expressed more generally, the number of coexisting species cannot be greater than the number of distinct regulating factors in the community \cite{levin1970community}.
For trophic communities of several levels, it was subsequently stated that the number of species on any trophic level could not exceed the sum of the numbers on adjacent levels \cite{levins1979coexistence}.
Experimental studies do demonstrate strong correlations between consumer and resource diversity \cite{sandom2013mammal, armbrecht2004enigmatic, andow1991vegetational,murdoch1972diversity}. 
These observations highlight that also the consumer plays a critical role in shaping the network of species, even when direct interaction between resource species is absent, an observation captured in Holt's paradigm of {\it apparent competition} \cite{holt1977predation}.

Despite the existing theoretical constraints and empirical findings, a selective theory for stable coexistence of many species in food webs is currently lacking.
This lack may partially be due to the complexity of the many-species interactions, yielding an uncontrollable number of parameters and hampering direct calculations or simulations of sufficient generality. 
Notwithstanding these complications, progress can be made when necessary conditions are demanded.
For an ecology, consisting only of a resource and a consumer level, we have recently shown that coexistence requires that the species richness of both levels is balanced and that a cascade of parameter values must be maintained \cite{haerter2014phage}. 
Examples of such systems may be the phage-bacteria ecology in the Atlantic Ocean \cite{moebus1983, flores2013multi} or laboratory ecologies.

However, in food webs, a subset of trophic levels can generally not be considered in isolation.
A species' niche is determined by its entire set of interactions, which generally may be composed of both beneficial and harmful interactions, i.e. the species may act both as a consumer or resource.
Further, many food webs contain omnivorous interactions, i.e. those where one species preys on several other species that are located at more than one trophic level.
To derive necessary conditions for coexistence in food webs, a more general starting point is required.

Based on the generalized Lotka-Volterra equation \cite{campbell1960}, we here show that in sustainable food webs each species must be part of a non-overlapping pairing. 
We define a non-overlapping pairing as a topological pattern for a directed network, where each species must contribute to a closed loop and none of the loops may overlap (Details: Methods).
Mathematically, this is a consequence of demanding that the determinant of the interaction matrix be nonzero.
We then analyze the implications of this pairing for the species richness at different trophic levels. 
We make predictions for secondary extinctions and assess the stabilizing effect of parasitism and omnivory.

\section*{METHODS}\label{sec:methods}
{\bf Steady state equations.} For consumer-resource interactions in food webs, the generalized Lotka-Volterra equations \cite{campbell1960} are
\begin{eqnarray}
 \dot{S}_i^{(1)}/{S_i^{(1)}} & = & k_i^{(1)} \left( 1-\sum_{j=1}^{n_1}p_{ji}S_j^{(1)}\right)-\alpha_i^{(1)}-\sum_{k=1}^{n_2}\eta_{ki}^{(2,1)}S_k^{(2)}\;
\label{eq:primaryProducer}
\end{eqnarray}
for primary producers and
\begin{eqnarray}
\dot{S}_k^{(l)}/S_k^{(l)} & = &  \sum_{m=1}^{n_{l-1}} \beta_{km}^{l,l-1} \cdot \eta_{km}^{l,l-1} \cdot S_m^{(l-1)}-\sum_{p=1}^{n_{l+1}} \eta_{pk}^{l+1,l} \cdot S_p^{(l+1)}- \alpha_k^{(l)}
\label{eq:higherLevel} 
\end{eqnarray} 
for species at trophic levels $l>1$.
We distinguish a species by the set of links that connect it to predators and prey or nutrients and the strength of these links (Details: Sec.~S4).
In Eqs~\ref{eq:primaryProducer} and~\ref{eq:higherLevel}, ${S_i}^{(l)}$ with $i=1, \dots ,n_l$ are the densities of species residing at trophic level $l$, $n_l$ is the species richness at level $l$, $k_i^{(1)}$ denote the maximal growth rates of $S_i^{(1)}$, $p_{ji}$ describe differential consumption of the basic resources by the $S_i^{(1)}$, $\alpha_i^{(l)}$ denote the decay rate of species $S_i^{(l)}$, $\eta_{ki}^{(l,l-1)}$ are the interaction coefficients between a species $S_k^{(l)}$ with species $S_i^{(l-1)}$, and $\beta_{km}^{l,l-1}$ are the efficiencies of reproduction of species $S_k^{(l)}$ when consuming species $S_m^{(l-1)}$.

In the steady state, the time derivatives $\dot{S}_i^{(1)}$ and $\dot{S}_k^{(l)}$ on the LHS of Eq.~\ref{eq:primaryProducer}, respectively Eq.~\ref{eq:higherLevel}, vanish and we have the equations
\begin{eqnarray}
 \sum_{j=1}^{n_1} p_{ji}S_j^{(1)}+\sum_{k=1}^{n_2}\frac{ \eta_{ki}^{(2,1)}}{k_i^{(1)}}S_k^{(2)}=\frac{k_i^{(1)}-\alpha_i^{(1)}}{k_i^{(1)}}&\equiv & \tilde{k_i}^{(1)}\;,\;\text{respectively} 
\label{eq:steady_state_primary} 
\\
\sum_{m=1}^{n_{l-1}} \beta_{km}^{(l,l-1)} \eta_{km}^{(l,l-1)} S_m^{(l-1)}-\sum_{p=1}^{n_{l+1}} \eta_{pk}^{(l+1,l)} S_p^{(l+1)} & \equiv & \alpha_k^{(l)}\;.
\label{eq:steady_state_higher}
\end{eqnarray}

Collecting all constant coefficients (RHS of Eqs \ref{eq:steady_state_primary} and \ref{eq:steady_state_higher}) in the vector $\bf{k}$ and all interaction coefficients on the LHS in the interaction matrix $\mathcal{R}$, we have the linear matrix equation ${\cal R} \cdot {\bf S} = {\bf k}$, where ${\bf S}$ is the vector of all species densities.
For completely shared nutrients, the competition factors $p_{ji}=1$. 
(Details: App. Secs S1 and S2).

\noindent
{\bf Parasite interactions.} Parasites have complex life-cycles that can demand several hosts \cite{huxham1996parasites,lafferty2008parasites}. Notwithstanding these complications, we here formally treat them as consumers, respectively resources of independently acting other species.
Also, we simplify concomitant links in terms of simple linear responses (details: SI Sec. S9.3.3).

\noindent
{\bf Non-overlapping pairing.} $det(\mathcal{R})\neq 0$ can be fulfilled if every species is paired with another species or nutrient (this constitutes a perfect matching \cite{lovasz1986matching}). 
For food webs with sharp trophic levels, the network is bipartite and therefore it is required that such a perfect matching exists. 
When species with variation in food chain length are present, one may generally obtain nonzero $det(\mathcal{R})$ by covering the entire network with {\it closed loops of directed pairings} (i.e. cycles).
This is a sequence of nonzero matrix elements $\mathcal{R}_{ij}$, $\mathcal{R}_{jk}$, $\mathcal{R}_{kl}, \dots ,\mathcal{R}_{mi}$ (Sec. S6), i.e. a chain of {\it directed pairings} where the direction is maintained and the last element connects to the first.
A directed pairing represents one nonzero matrix element, whereas a pairing also includes the symmetric element. 

\noindent
{\bf Empirical food web data.} 
We use high-resolution data on seven food webs including free-living and parasite species: 
The North American Pacific Coast webs Carpinteria Salt Marsh (CA), Estero de Punta Banda (PB), Bahia Falsa (BF) \cite{hechinger2011food, lafferty2006parasites};  the coastal webs Flensburg Fjord (FF) \cite{zander2011food}, Sylt Tidal Basin (ST) \cite{thieltges2011food}, and Otago Harbor (OH), New Zealand \cite{mouritsen2011food}, as well as the Ythan Estuary (YT), Scotland \cite{huxham1996parasites}. These food webs describe consumer-resource interactions between basal, predatory and parasite species.
A compilation of all seven food webs has recently been provided \cite{dunne2013parasites, dunnedata:2013}. 
Specifically, the data distinguish three types of links: (i) links between free-living species only (``Free''), (ii) additional links between parasites and other species (``Par'') and (iii) links from free-living consumers to the parasites of their resources (``ParCon''), i.e. so-called concomitant links.
Details on data analysis: SI Section S9.
For the empirical data, the lack of niches, i.e. nullity $d\equiv S-\text{rank}(\mathcal R)$, was computed by using random values for all nonzero entries of the respective matrix ${\mathcal R}$. 
Basal species were each given an individual nutrient source. In ``ParCon sym'' a subset (20 percent) of concomitant links were randomly selected to be symmetric (Details: Sec. S9.3.3).
In the data analysis and simulations (Fig.~\ref{fig:emp_interaction_matrix} and \ref{fig:sim_Matrices}), the trophic level of a species is defined by its prey-averaged food chain length (Sec. S9.2.1). 

\noindent{\bf Simulations.} 
We perform two types of simulations:
(i) In-silico assembly of a tree-like food web (Fig.~\ref{fig:interaction_matrix}c), where parameters are chosen according to constraints discussed in SI Sec.~S2.
The numerical values of the parameters are: For the interaction and growth coefficients $\eta=\beta=k=1$ for all links present (solid black arrows in Fig.~\ref{fig:interaction_matrix}b), as well as the decay coefficients $\{\alpha_1,\dots,\alpha_{8}\}=\{.1,.1,.16,.1,.12,.15,.1,.1\}$, where the labels are as indicated in Fig.~\ref{fig:interaction_matrix}b.
Each new species is introduced at low density and time-integration is continued until steady state is reached (using Mathematica NDSolve method).
(ii) An idealized food web was constructed by using the average species counts at levels $n_i$ obtained from all empirical data sets and initially assuming sharp trophic levels for all species. 
Sharp trophic levels were obtained by rounding each species' chain length to the nearest integer value (Fig.~\ref{fig:sim_Matrices}a).
With the constraint of these trophic levels, a number of links was assigned to match the empirical average for free-living food webs (Fig.~\ref{fig:sim_Matrices}b).
When adding further species, the empirical average of parasite species count was used (47 species). 
To obtain Fig.~\ref{fig:sim_Matrices}, initially, each parasite received one link.
In the cases (c) and (d), this link connected the parasite to any existing free-living species.
In the cases (e) and (f), this link connected the parasite to any existing species at trophic levels 3 or 4.
For any subsequent link, a parasite was chosen at random. 
A link was then formed in three ways: 
Case c: randomly to connect with another existing species; 
Case d: randomly, but only to species at the same trophic level as for the initial link of that parasite;
Case e: randomly to other existing species at levels 3, level 4;
Case f: randomly to other existing species at levels 3, level 4 or another parasite (More detail: Sec. S10).

\section*{RESULTS}\label{sec:results}
\subsection*{Theory}\label{sec:theory}
We describe the interaction of species on $L$ trophic levels by the generalized Lotka-Volterra equations \cite{campbell1960,case2000illustrative}. Basal species are constrained by the system carrying capacity while the consumers are assumed not self-limiting, and trophic interactions occur through the linear type-I functional response (Methods). 
Such equations have been widely used in community assembly models, where food web networks are assembled by numerically analyzing the equations to find parameter sets with stable and/or permanent coexistence solutions 
\cite{post1983community, law1996permanence, drossel2002modelling}. 
Here we take an alternative path by first finding a {\it necessary condition} for coexistence in terms of species richness, which results in the food web assembly rules that constrains the network topology.
We subsequently show that when these assembly rules are fulfilled, stable network structures can always be obtained.

In the steady state, we have the matrix equation ${\cal R} \cdot {\bf S} = {\bf k}$, where ${\bf S}$ is the vector of all species densities, ${\cal R}$ is the interaction matrix between the species, and ${\bf k}$ is the vector of growth and decay coefficients.
Note that $\mathcal{R}$ has a block structure with nonzero entries only for interactions between neighboring trophic levels but not within the same level, and that the positions of these matrix elements are symmetric due to mutual interaction between predator and prey (Fig.~\ref{fig:interaction_matrix_general}a).
Stable/permanent coexistence requires that a feasible solution ${\bf S^*}\equiv{\cal R}^{-1}{\bf k}> 0$ exists \cite{allesina2012stability,law1996permanence}. Structural stability, i.e. robustness against parameter perturbations, of a feasible coexistence solution requires that $det({\cal R} )$ be nonzero (Sec.~S2),  a condition required for the existence of the matrix inverse in a linear equation \cite{petersen1962linear}.
In the following, we analyze what this basic condition means for the species richnesses at the different trophic levels of a food web.
We then discuss also the stability of a given solution.

We first specialize to the case of a single shared basic nutrient, e.g. sunlight, used by all primary producers $S_i^{(1)}$, with $n_1$ the species richness on the first trophic level and $i$ ranging from $1$ to $n_1$. 
This corresponds to setting the competition terms $p_{ji}=1$ for all $1\leq i$,$j\leq{n_1}$ in Eq.~\ref{eq:primaryProducer}, thereby yielding a block of $n_1\times n_1$ unit entries in the lower right block. 
(Example: Sec.~S4).
Nonzero determinant is achieved if it is possible to identify a path 
of matrix elements that only contains elements from the non-zero sub-matrices bordering the diagonal (Fig.~\ref{fig:interaction_matrix_general}a). 
This is equivalent to demanding that every species be part of a consumer-resource pair connecting neighboring trophic levels and none of these pairs overlap (known as {\it perfect matching} of a bipartite graph in graph theory \cite{lovasz1986matching}).
The pairing guarantees that no species share exactly the same niche, i.e. a particular set of interactions with resources and consumers (Sec. S4), and manifests the competitive exclusion principle (Fig.~\ref{fig:interaction_matrix_general}b, i and ii).
For primary producers, pairings may involve the nutrient source (Fig.~ \ref{fig:interaction_matrix_general}a, inset).
In that case, at least $n_1-1$ species at level two are required for pairing of the remaining basal species.  
We term this structure {\it resource-limited}. 
In this case, $n_2\geq n_1-1$ is required. 
In the example (Fig.~ \ref{fig:interaction_matrix_general}a) this condition can indeed be fulfilled, because a sufficient number of species exists at level two.
If pairing with nutrients is not used,  
$n_2\geq n_1$ is required --- a {\it consumer-limited} configuration due to the biomass restriction imposed by consumer predation (Fig.~S2). 
In turn, those species left unpaired at level two must be paired by species at level three.

Defining $N_o$ and $N_e$ as the sums of node richness at odd and even levels, respectively, 
it is then easy to check that the general constraints become
\begin{equation}
 \Delta\equiv N_o-N_e\in \{0,1\}\;,
 \label{eq:N_r_N_c_v1}
\end{equation}
which encompasses the competitive exclusion principle \cite{gause1936struggle,hardin1960competitive,haerter2014phage}. 
Defining $L$ as the top trophic level, for each of the two options in Eq.~\ref{eq:N_r_N_c_v1} a set of $L-1$ nested inequalities arises relating the species counts $n_i$:
\begin{equation}
\begin{aligned}
 (i)\\
 (ii)\\
 (iii)\\	
 (iv) \\
 \\
\end{aligned}
\hspace{1cm}
\begin{aligned}
n_1&\geq & \Delta\\
n_2 & \geq & n_1-\Delta \\
n_3 & \geq & n_2-n_1+\Delta \\ 
n_4 & \geq & n_3-n_2+n_1-\Delta \\
\vdots
\end{aligned}
\label{eq:examples}
\end{equation}

\noindent
This sequence continues until $n_L$ is reached.
The case $\Delta=0$ is consumer-limited, whereas $\Delta=1$ signals resource limitation.
Note that these rules allow simple assessment on which food webs can have coexistence-solutions.
To illustrate this, we give two simple examples of food webs that cannot coexist by Eqs~\ref{eq:N_r_N_c_v1},\ref{eq:examples} (Fig.~\ref{fig:interaction_matrix_general}b, iii and iv). 
Our rules thereby are more selective than those in previous work \cite{levins1979coexistence}. 
There, a requirement was stated for trophic communities of several levels, where the number of species on any level could not exceed the sum of the numbers on adjacent levels --- a condition that would e.g. {\it not} rule out the two simple webs (Fig.~\ref{fig:interaction_matrix_general}b, iii and iv), hence not detect the lack of niches.

The conditions (Eqs~\ref{eq:N_r_N_c_v1},\ref{eq:examples}) rule out stable coexistence when violated, yet they leave unanswered if all populations can be positive or stable when they are met.
For any choice of species counts fulfilling Eqs~\ref{eq:N_r_N_c_v1},\ref{eq:examples}, a feasible solution can be obtained by starting from the complete non-overlapping pairing of species, i.e. a subset of links. 
A tree-like backbone of links will be obtained, which preserves the complete non-overlapping pairing. 
Additional links give each species access to nutrients.
For this structure, we find that the parameters can always be assigned to yield positive steady-state population densities for all species. 
This is possible by enabling trade-offs between the invader and the resistant species, so that the fitness of any invader, e.g. described by its decay coefficient, is sufficiently large to maintain competitiveness of the residents \cite{haerter2015inprep}.
An example is shown in the subsequent section.

The existence of a Lyapunov function is a sufficient criterion for global stability \cite{goh1977global}. 
Using this, in extension of the analysis of simple chains of species \cite{hofbauer1988}, we demonstrate that these food webs are always stable, i.e. the Lyapunov function decreases monotonically with time after the initial perturbation (Sec.~S2.2).
We thereby establish that for each combination of species richnesses consistent with the conditions (Eqs~\ref{eq:N_r_N_c_v1},\ref{eq:examples}), a feasible and stable food web exists. 

In practice, there may be several basic nutrients, such as different chemical compounds or spatial or temporal subdivision \cite{tilman1994competition}.
If $n_S>1$ separate nutrients are available, the assembly rules yield $n_S$+$1$ sets of conditions analogous to those in Eq.~\ref{eq:examples} where any $\Delta\in \{0,n_S\}$ is allowed (Sec.~S7).
The presented theory assumes simple predator-prey couplings. 
Non-linear interactions (i.e. type-II response) and cannibalism can be included by adding diagonal matrix elements in Fig.~\ref{fig:interaction_matrix_general}a, corresponding to species that pair with themselves (Secs~S7,S8). 
Another future extension of the theory could be that of incorporating frequency-dependent predation \cite{allen1988frequency}.

\subsection*{Food web assembly}\label{sec:assembly}
We return to a single nutrient ($n_S=1$) and now discuss species richness at the different trophic levels. 
In the consumer-limited case, 
Eq.~\ref{eq:examples}$(i)$ restricts $n_1$ to an upper bound given by $n_2$. 
Eq.~\ref{eq:examples}$(ii)$ conversely restricts $n_2$ to numbers equal to the sum of the total species richness in both neighboring trophic levels, hence the limitation to the abundance of $n_2$ is much weaker compared to that of $n_1$.
Accordingly, the basal level cannot constitute the global maximum of node richness within a level. 
Moving further, $n_3$ could again exceed $n_2$ but possible limitations stemming from the count of top predator species become noticeable.
When starting at the top predator level $L$, by symmetry, similar constraints as in Eq.~\ref{eq:examples} hold: $n_{L}-n_{L-1}\leq 0$.
Together with Eq.~\ref{eq:examples}$(i)$ we have for the consumer-limited state, $n_1+n_{L}\leq n_2+n_{L-1}$, i.e. species at intermediate trophic levels generally dominate food web biodiversity (Fig.~\ref{fig:trophic_cartoon}). 
The condition is similar for the resource-limited state, where the limit is shifted by one species.
For increasing $L$, these equations predict further increase in the contribution of intermediate species to total biodiversity. 

Given these general constraints, we can now discuss food web assembly. 
Consider a simple food web and its interaction matrix (Fig.~\ref{fig:interaction_matrix}a). 
Graphically, an allowed structure is again manifested by permitting non-overlapping pairing of species.
Food web growth is characterized by alternating transitions between resource- and consumer-limited states. 
Initial community growth requires the presence of a single primary producer (Fig.~\ref{fig:interaction_matrix}b), hence $n_1=1$.
The only possible addition is then a consumer, preying on the existing producer.
This entails immediate addition of a trophic level ($n_1=n_2=1$) and transition to a consumer-limited state.
The assembly rules subsequently permit additions at different trophic levels, but alternation between consumer and resource limitation must be preserved.

For each food web consistent with the conditions in Eqs~\ref{eq:N_r_N_c_v1},\ref{eq:examples}, parameters can be assigned in a way that food web evolution is possible, i.e. that feasibility and stability are even achieved after each addition of species. 
To exemplify this, we assign parameters for all eight species of the example, tree-like, food web (Fig.~\ref{fig:interaction_matrix}b, \cite{haerter2015inprep}).
Using in-silico simulations, we mimic the evolution of this web, starting from only a single species (Fig.~\ref{fig:interaction_matrix}c). 
Any addition of new species leads to a transient disruption of population densities, but a new feasible steady state is eventually established (Fig.~\ref{fig:interaction_matrix}c). 
To quantify stability, we further compute a Lyapunov function, and verify that it monotonically decreases after each addition within each stage of the food web (Sec. S2). 
We have checked that addition of further weak links is compatible with stability and feasibility, hence in principle allowing arbitrary link structure.

Our assembly rules are easily generalized to food webs containing omnivores or parasites with hosts at several trophic levels (Sec.~S6).  
Consider again Fig.~\ref{fig:interaction_matrix}a, but imagine that another species is added on trophic level 4, causing a violation of the assembly rules.
This violation can be rectified by a single generalist omnivore (Fig.~\ref{fig:interaction_matrix_omnivore}), corresponding to an additional row and column of nonzero coefficients in the interaction matrix.
The omnivore can be interpreted as a consumer preying on all trophic levels. 
Hence, one can choose a level $i$ and let its species count $n_i$ be increased by one unit to again satisfy the assembly rules.
In general, omnivory or parasitism can make the graph non-bipartite. If so, the non-overlapping pairing to achieve $det(\mathcal{R})\ne 0$ is extended to covering the entire network with closed loops of directed pairings (Sec.~S6). 

\subsection*{Data analysis and simulations}\label{sec:data_analysis}
What do the assembly rules teach us about real food webs? 
For seven detailed empirical food webs (Details: Materials and Methods) containing both free-living and parasite species, we determine the difference between the respective total number of species $S$ and the maximum number of linearly independent rows, $d\equiv S-\text{rank}(\mathcal R)$ (Details: Methods). 
Linear dependence can be seen as the sharing of a specific niche by several species, therefore $d$ measures the lack of niches in the given food web. 
When $d=0$, the assembly rules are satisfied ($det(\mathcal R)\ne 0$). 
As done previously \cite{dunne2013parasites}, we distinguish webs formed by: links between free-living species only; links between all free-living and parasite species; with additional concomitant links (Methods and SI Sec. S9.3.3).

The free-living webs have substantial structure, an example is given in Fig.~\ref{fig:emp_interaction_matrix}a and others in Sec.~S9. Many species are located at sharp trophic levels (Fig.~S7), a feature that manifests itself by the blocks of white spaces, i.e. absent interactions (Fig.~\ref{fig:emp_interaction_matrix}a).
The organization into sharp trophic levels entails stricter demands on the combinations of species richnesses (Eqs~\ref{eq:N_r_N_c_v1},\ref{eq:examples}), and many sub-webs consisting only of free-living species do not seem to fulfill these demands. 
This is quantified by an associated lack of niches in most free-living webs, i.e. $d>0$. 
Additional analysis reveals that all free-living webs are in a consumer-limited state, i.e. species richness in even trophic levels dominates (Sec. S9).

We contrast these findings with simple models, namely the cascade \cite{cohen1985stochastic} and niche models \cite{williams2000simple}.
Using number of species and links from the empirical datasets, we generate network samples (Fig.~\ref{fig:emp_interaction_matrix}b,c). 
The resulting interaction matrices are characterized by very little structure in terms of trophic levels (the white blocks are all but missing).
When repeating the simulations for all seven webs (Sec.~S9) and obtaining the corresponding rank deficiencies, we find that the networks simulated using the cascade and niche models consistently give $d\approx 0$ and are less structured than empirical data (Sec.~S9).

We further quantify the organization of species by the chain length distributions for the empirical and modeled networks, where much broader distributions are found for the models.
Quantifying omnivory by the standard deviation of chain lengths \cite{lafferty1996altered}, modeled networks consistently yield substantially higher fractions of omnivory.

We now consider the webs involving parasites (Fig.~\ref{fig:emp_interaction_matrix}d,e). 
At the edge of the panels we indicate by a color-coding, where, in terms of trophic level, parasites enter and how the free-living species are re-organized. 
Notably, parasites predominantly enter at high trophic levels, the lower section (approximately levels one and two, red to green colors) remains nearly unaltered by the inclusion of parasites.
Specifically, interactions of a given parasite generally involve several free-living species at multiple trophic levels (Sec.~S9), acting to loosen the structure at the higher levels ($n_3$ and $n_4$, compare Fig.~\ref{fig:emp_interaction_matrix}a,d) and reaching agreement with the assembly rules (Sec.~S6).
In other words, some of the parasites can be seen as effectively acting as odd-level species, thereby relaxing the initially consumer-dominated free-living webs to a more balanced state.

Concomitant links (Fig.~\ref{fig:emp_interaction_matrix}e) cause further entanglement of trophic levels, open additional options for possible pairings and systematically increase sustainability of food webs. 
Concomitant links require additional consideration, as they generally are directed links where a parasite is consumed by its host's predator, i.e. they denote a detrimental effect on the parasite population.
A positive impact on the predator's population may however not always result.
We call such links {\it asymmetric concomitant links}.
If the predator's population does benefit from the consumption, such as observed in some studies \cite{lafferty1996altered,johnson2010parasites}, we use the term {\it symmetric concomitant link}.
Nonetheless, such directed links can lead to additional non-overlapping pairings, when a closed loop of directed links is formed, e.g. a triangle (loop of length three).
In the empirical webs it is noticeable that inclusion of asymmetric concomitant links only rarely yields rank improvement. 
Investigating this further, we find that many food webs with parasite interactions already contain sufficient numbers of loops to allow pairings between parasites and free-living species. 
The limitation arises because a surplus of parasites exists. 
Each loop will involve at least two free-living species but only one parasite, making it impossible to find non-overlapping pairings for all parasites.
For those webs, only the inclusion of symmetric concomitant links leads to an additional improvement of rank, since then each parasite can be paired with a single free-living species.
In two webs, where loops are rare, even asymmetric concomitant links help improve the rank.
Details and simulations are available in SI sections S9.3.3 and S10.4.

For all empirical food webs, we summarize the effect of the different link additions on rank deficiency $d$ (Fig.~\ref{fig:emp_interaction_matrix}f). 
We find a general decrease of $d$ as more parasite links are added.
Notably, full rank is sometimes not achieved, even when all available concomitant links are included. 
Indeed, empirical food web datasets may often be incomplete, as some links can be difficult to detect.
Our findings may serve as a means of identifying possibly missing data, most notably in the Ythan Estuary food web, where overall link density is low and parasite-parasite links are completely absent (further details on individual food webs: Sec.~S9.3.2 and S10.3).

As mentioned above, existing food web models generally produce network structures lacking rank deficiency ($d\approx 0$), even for the free-living webs.
When instead starting from model networks with similar link density as the empirical webs but sharp trophic levels, we obtain $d$ substantially larger than zero (Fig.~\ref{fig:sim_Matrices}).
Parasites, with their complex life-cycles \cite{huxham1996parasites}, often consume species at varying trophic levels during different stages of their lives.
Adding species that each interact with species at multiple trophic levels (Fig.~\ref{fig:sim_Matrices}c), i.e. mimicking the addition of parasites, 
our modeled webs give systematic decrease of $d$ as links are added (black line in Fig.~\ref{fig:sim_Matrices}g).
When adding species that each interact with a single (Fig.~\ref{fig:sim_Matrices}d), or exclusively higher trophic levels (Fig.~\ref{fig:sim_Matrices}e), saturation at $d>0$ occurs (blue and red lines in Fig.~\ref{fig:sim_Matrices}b).
When simulating also parasite-parasite interaction, $d$ is also found to decrease (Fig.~\ref{fig:sim_Matrices}f, Simulation details: Methods and Sec. S10).
Overall, these simulations suggest that addition of random species without the feature of interaction with multiple trophic levels is not sufficient to explain removal of rank deficiency.

We have further explored addition of omnivorous links to the free-living web. 
The rank deficiency $d$ is reduced rapidly if the addition happens randomly for all trophic levels, but the reduction is limited if omnivorous links occur mostly at the trophic level 3 as in the real data (Sec. S10.2).   We have also performed extensive simulations on the effect of concomitant predation, which further emphasizes the importance of parasite-parasite interaction in achieving coexistence for some webs (Sec. S10.4).


\section*{DISCUSSION}\label{sec:discussion}
The food web assembly rules generalize the competitive exclusion principle to food webs of any number of species and trophic levels.
They quantify which combinations of species richness at the different trophic levels can yield coexistence solutions. 
We show that for any of these combinations, there are stable and feasible network structures. 
Demanding full rank of the food web interaction matrix expresses the simple notion that each species must occupy a unique niche 
and leads to biologically plausible combinations of species richnesses at the different trophic levels.
The requirement, i.e. non-overlapping pairing or equivalently nonzero determinant, is simple and directly allows us to evaluate the self-consistency of empirical data.
The rules help explain that actual food web networks are far from random and more structured than those obtained from the traditional niche and cascade models. 
While some food web datasets do fulfill our conditions, networks known to lack interactions, e.g. the Ythan Estuary web \cite{huxham1996parasites}, stand out as particularly far from reaching agreement with the rules.

One immediate consequence for food webs with species predominantly organized according to trophic levels, e.g. many free-living webs, is that species richness at the basal and top-predator levels should be limited by the species richness of the respective neighboring levels (compare Eq.~\ref{eq:examples}).
This can explain the observations, both for terrestrial \cite{Martinez1991} and marine food webs \cite{dunne2004network}, which report greatest species richness at intermediate trophic levels while top predators and basal species contribute little. 
Similar conclusions were further drawn from semi-analytical work \cite{bastolla2005biodiversity2}, where a maximum in biodiversity at an intermediate trophic level was predicted.
Another consequence for such webs is that additions of species are generally not possible at any trophic level, if sustainable ecologies are to be achieved.
Even when the addition satisfies the assembly rules, its presence might cause substantial redistribution of biomass, i.e. shifts between consumer and resource limitation. 
In practice, it may be precisely these dramatic transitions that explain the profound and abrupt impacts on species abundance and energy flow patterns which are sometimes observed in the field. 
E.g. the introduction of opossum shrimp into a lake caused a cascade of trophic disruptions by reduction of salmon numbers and subsequent depletion of eagle and grizzly bear \cite{spencer1991shrimp}.
On the other hand, our rules also describe the circumstances, under which removal of a species must trigger additional extinctions.

The assembly rules thus allow predictions of secondary extinction, resulting either from addition or removal of species.
If the modified food web obeys the assembly rules, the food web might be stable. 
Indeed, in some observed cases, ecological release of new species into a habitat has had relatively gentle effects \cite{williamson1996varying}. 
However, a violation of the assembly rules (Eqs~\ref{eq:N_r_N_c_v1},\ref{eq:examples}) by addition of a new species can have one of two effects: 
Either the new species will not be competitive and collapse, or a number of species will collapse (possibly including the species itself) to restore the food web to a permitted state.
For removal of a species that leads to violation of the assembly rules, secondary extinctions \cite{eklof2006species} must be triggered to re-gain a sustainable state.
We find a consistent pattern, when considering species removal in empirical food webs: 
E.g. in consumer-limited webs, such as the free-living empirical webs, secondary extinctions are more likely triggered by removal of resource than consumer species (Sec. S9 and Fig.~S12).

Community omnivory \cite{fagan1997omnivory,mccann2000diversity,bascompte2005interaction,thompson2007trophic} and parasitism \cite{hudson2006healthy,lafferty2008parasites, dunne2013parasites} have been suggested as contributing to food web stability. 
Our approach provides theoretical support for this claim.
We indeed find that full rank of the food web interaction matrix is difficult to achieve for species that are organized at strict trophic positions. 
Species that consume resources at different trophic positions, e.g. omnivores or some parasites, are shown to loosen the constraints and make it easier to comply with the assembly rules, i.e. finding a non-overlapping pairing of species.

\pagebreak
\section*{ACKNOWLEDGMENTS}
We thank I. Dodd and S. Semsey for fruitful discussions.
The authors acknowledge financial support by the Danish National Research Foundation through the Center for Models of Life.

\section*{Competing Interest Statement}
The authors have no competing interests to declare.





\setcounter{figure}{0}


\setcounter{figure}{0}
\begin{center}
\begin{figure*}
\begin{center}
\begin{overpic}[width=5cm,angle=-90,trim= 0cm 0pt 0pt 0pt,clip]{dummy.eps}
\put(-32,-10){\includegraphics[width=7.4cm,angle=0,trim= 0cm 3.7cm 0pt 0pt,clip]{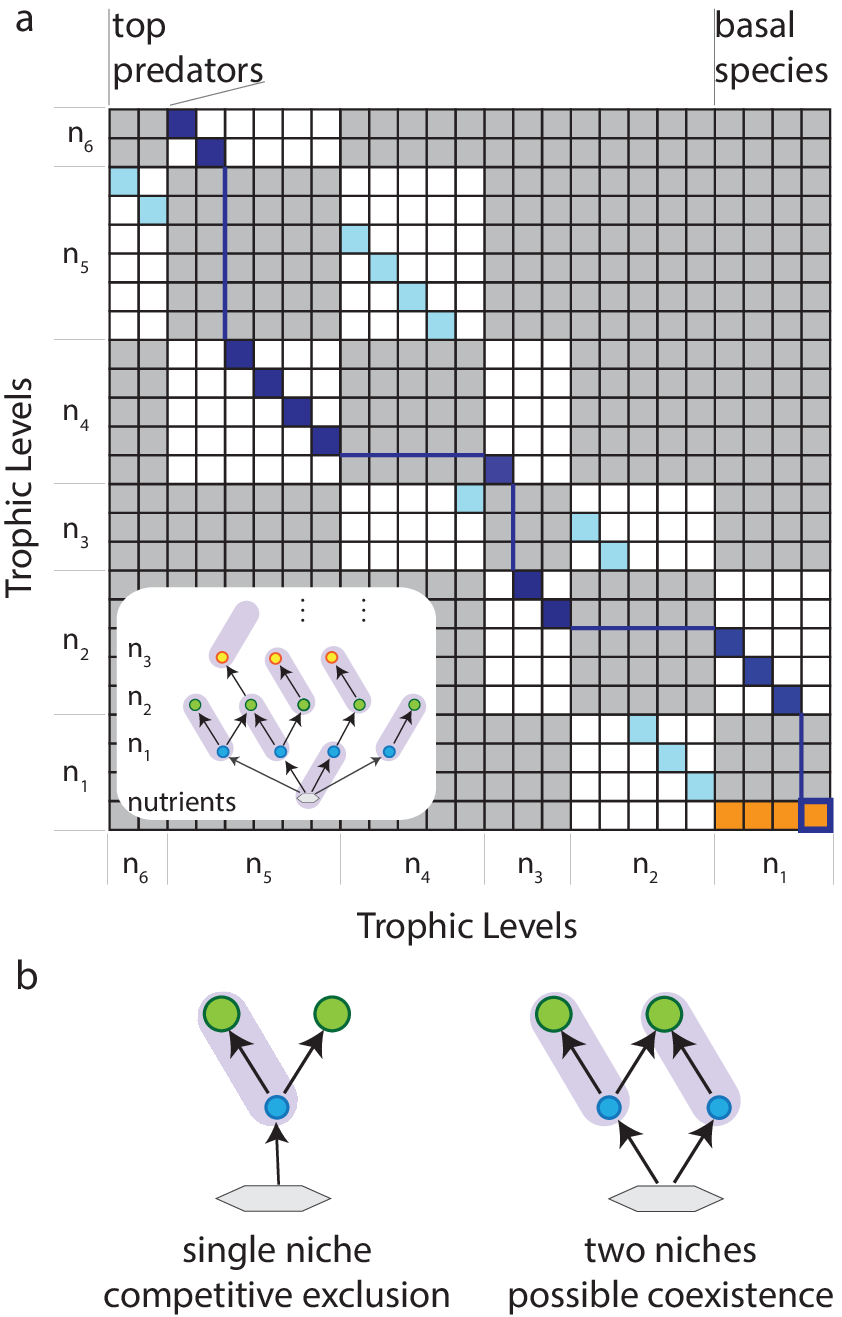}}
\put(-32,-90){\includegraphics[width=7.4cm,angle=0]{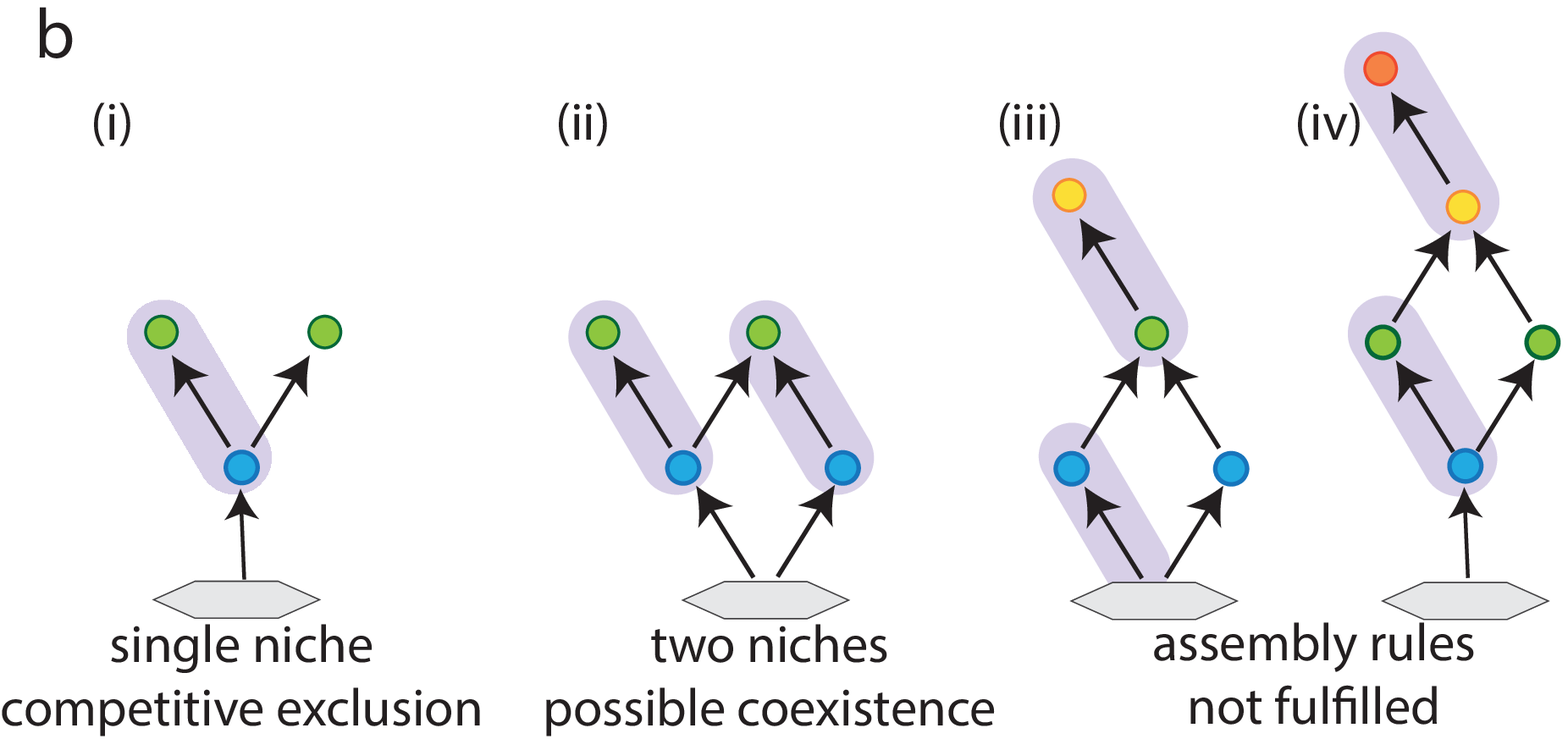}}
\end{overpic}
\vspace*{140pt}
\caption{\doublespacing{\bf Food web interaction matrix and application of perfect matching.}  
{\bf a}, 
White (gray) boxes indicate nonzero (zero) matrix elements, orange boxes are unity matrix elements for the primary producers; dark and light blue squares indicate a possible path chosen, allowing $det({\mathcal R})$ to be nonzero. 
Here, $N_o=n_1+n_3+n_5=13$ and $N_e=n_2+n_4+n_6=12$, and Eq.~\ref{eq:examples} is fulfilled with $\Delta=1$.
Inset: Schematic of a possible pairing for the chosen path.
Note that the invariance property of $det({\mathcal R})$ was used, yielding only $n_1$ non-vanishing matrix elements in the lower right block (Details: SI). 
{\bf b}, Perfect matching \cite{lovasz1986matching} applied to simple food webs where competitive exclusion rules out coexistence due to lack of niches (i) and where enough niches are available for coexistence (ii). (iii) and (iv) are two additional examples, where coexistence is ruled out by the assembly rules. 
In (iii), $n_1$=$2$, $n_2$=$n_3$=$1$. In (iv), $n_1$=$n_3$=$n_4$=$1$, $n_2$=$2$. In both, $\Delta\equiv N_o-N_e\notin \{0,1\}$, see Eq.~\ref{eq:N_r_N_c_v1}.
}
\label{fig:interaction_matrix_general}
\end{center}
\end{figure*}
\end{center}

\begin{center}
\begin{figure*}
\begin{center}
\begin{overpic}[width=5cm,angle=-90,trim= 0cm 0pt 0pt 0pt,clip]{dummy.eps}
\put(-15,-80){\includegraphics[width=5.4cm,angle=0]{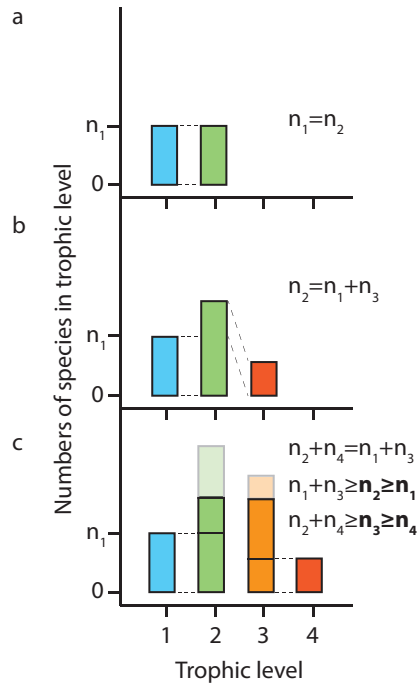}}
\end{overpic}
\vspace*{140pt}
\caption{\doublespacing{\bf Species richness for different trophic levels.}  
{\bf a}, Food web with two trophic levels only; a staircase of coexistence with balanced species richness at levels 1 and 2 \cite{haerter2014phage}. 
{\bf b}, Three trophic levels. The number of intermediate species must equal the total number of basal and predator species. Intermediate species dominate ecosystem biodiversity. 
{\bf c}, Four trophic levels. $n_2$ ($n_3$) must at least match basal (predator) species richness $n_1$ ($n_4$), indicated by thin black lines in green and orange bar. Solid green (orange) bars show the minimal upper bound to species richness in trophic level one (two). 
Species richness $n_2$ and $n_3$ can increase even further by co-evolution of intermediate species (shaded region). 
Note the applicable assembly rules shown for the different cases. }
\label{fig:trophic_cartoon}
\end{center}
\end{figure*}
\end{center}

\begin{center}
\begin{figure*}
\begin{center}
\begin{overpic}[width=5cm,angle=-90,trim= 0cm 0pt 0pt 0pt,clip]{dummy.eps}
\put(-90.5,-15){\includegraphics[width=13.45cm,angle=0]{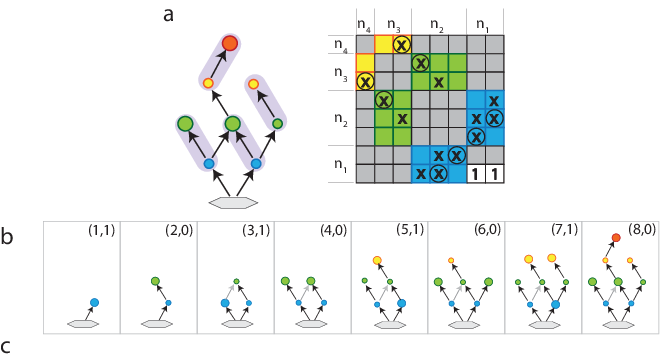}}
\put(-85.5,-65){\includegraphics[height=2.9cm,angle=0,trim= 0cm 0cm 0pt 0pt,clip]{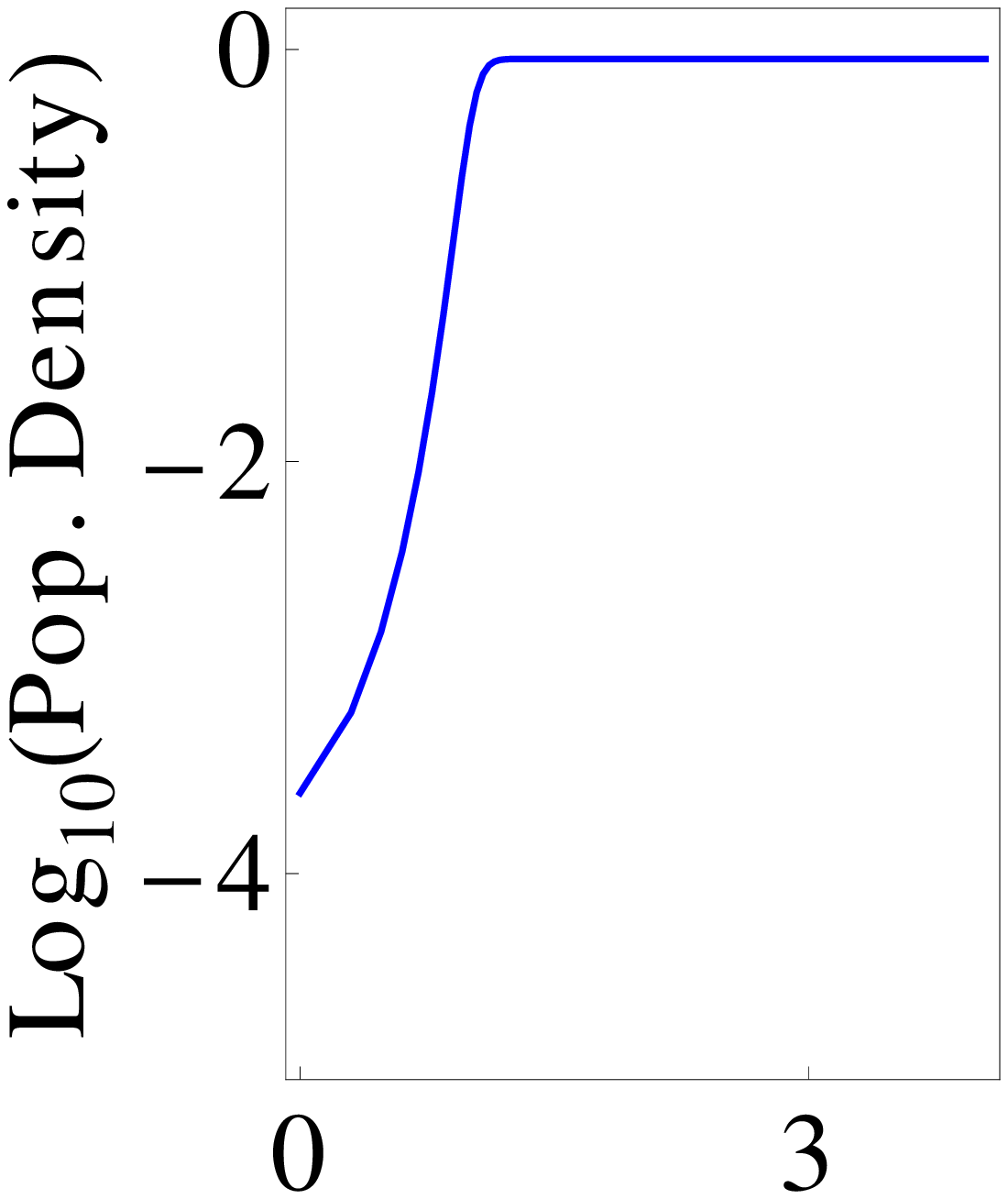}}
\put(-41.5,-65){\includegraphics[height=2.9cm,angle=0,trim= 3.2cm 0cm 0pt 0pt,clip]{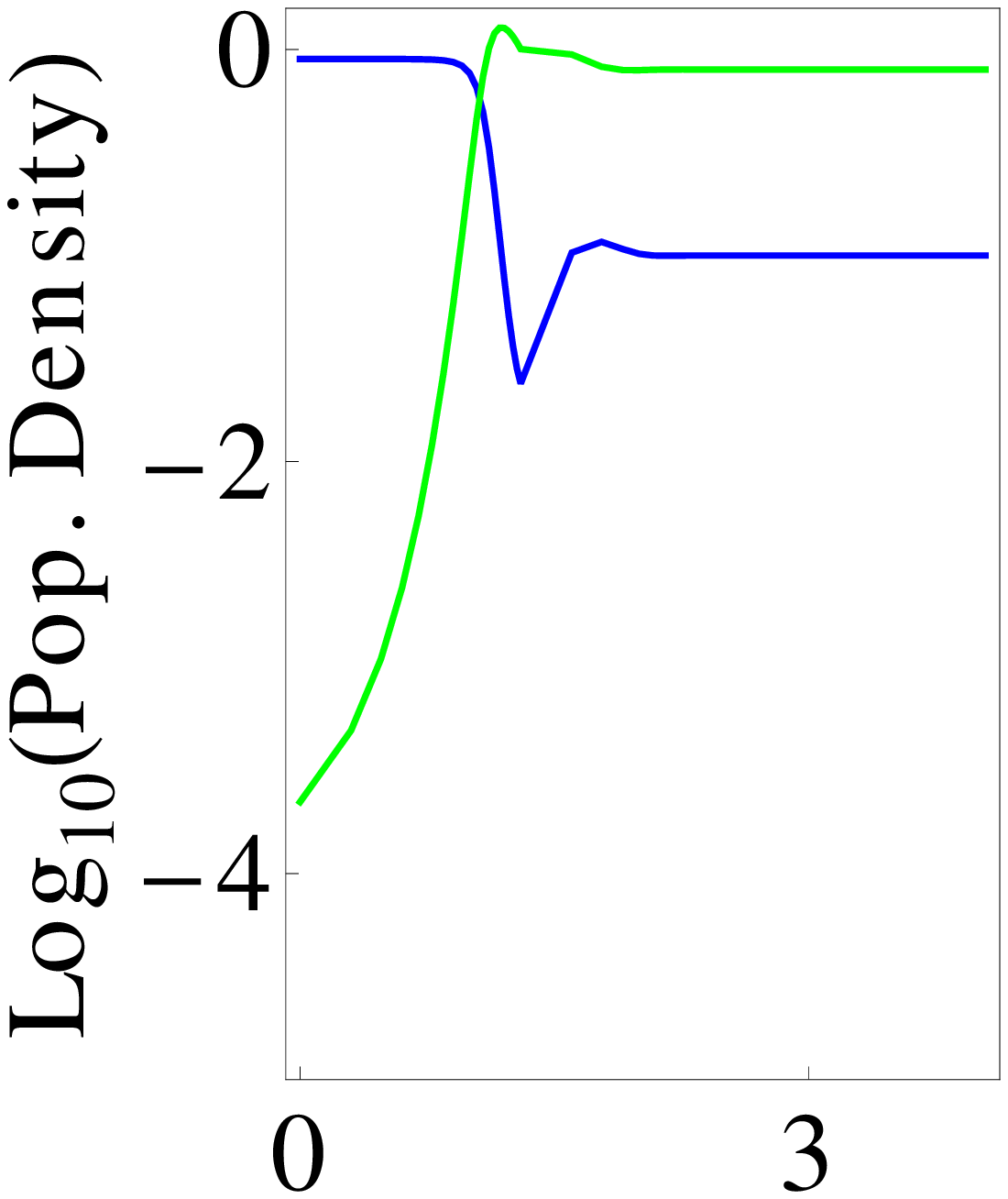}}
\put(-10.25,-65){\includegraphics[height=2.9cm,angle=0,trim= 3.2cm 0cm 0pt 0pt,clip]{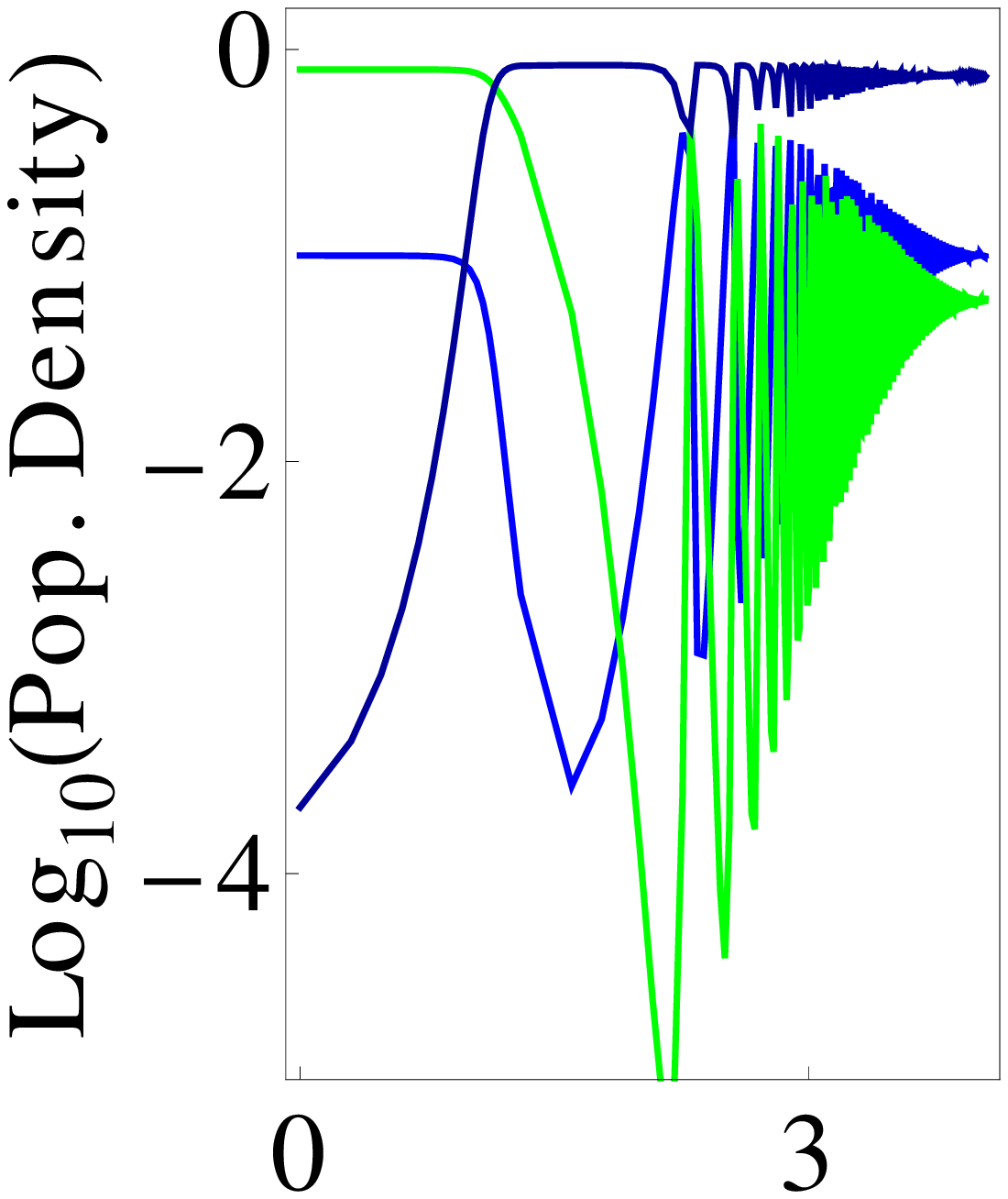}}
\put(21.5,-65){\includegraphics[height=2.9cm,angle=0,trim= 3.2cm 0cm 0pt 0pt,clip]{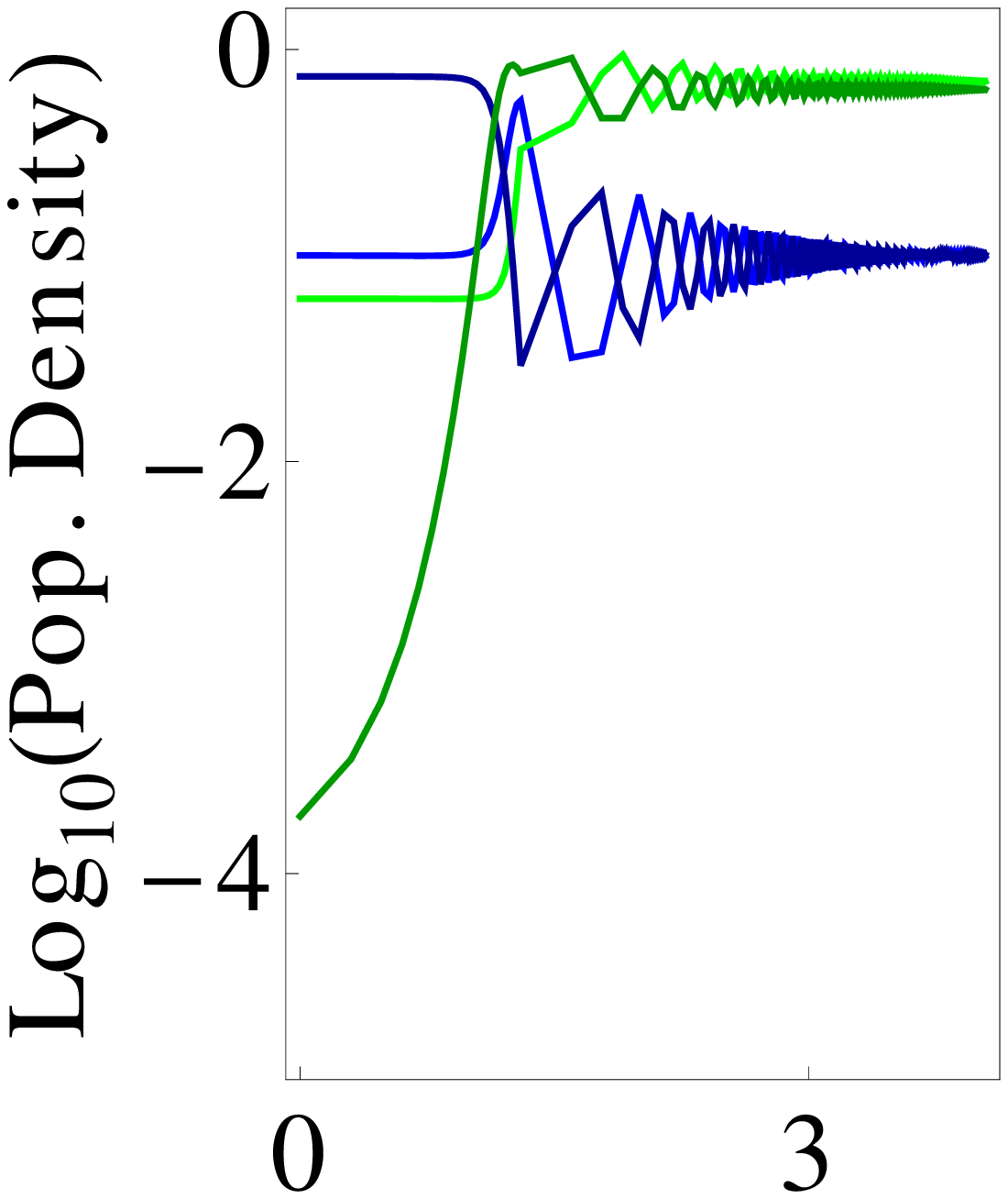}}
\put( 52.75,-65){\includegraphics[height=2.9cm,angle=0,trim= 3.2cm 0cm 0pt 0pt,clip]{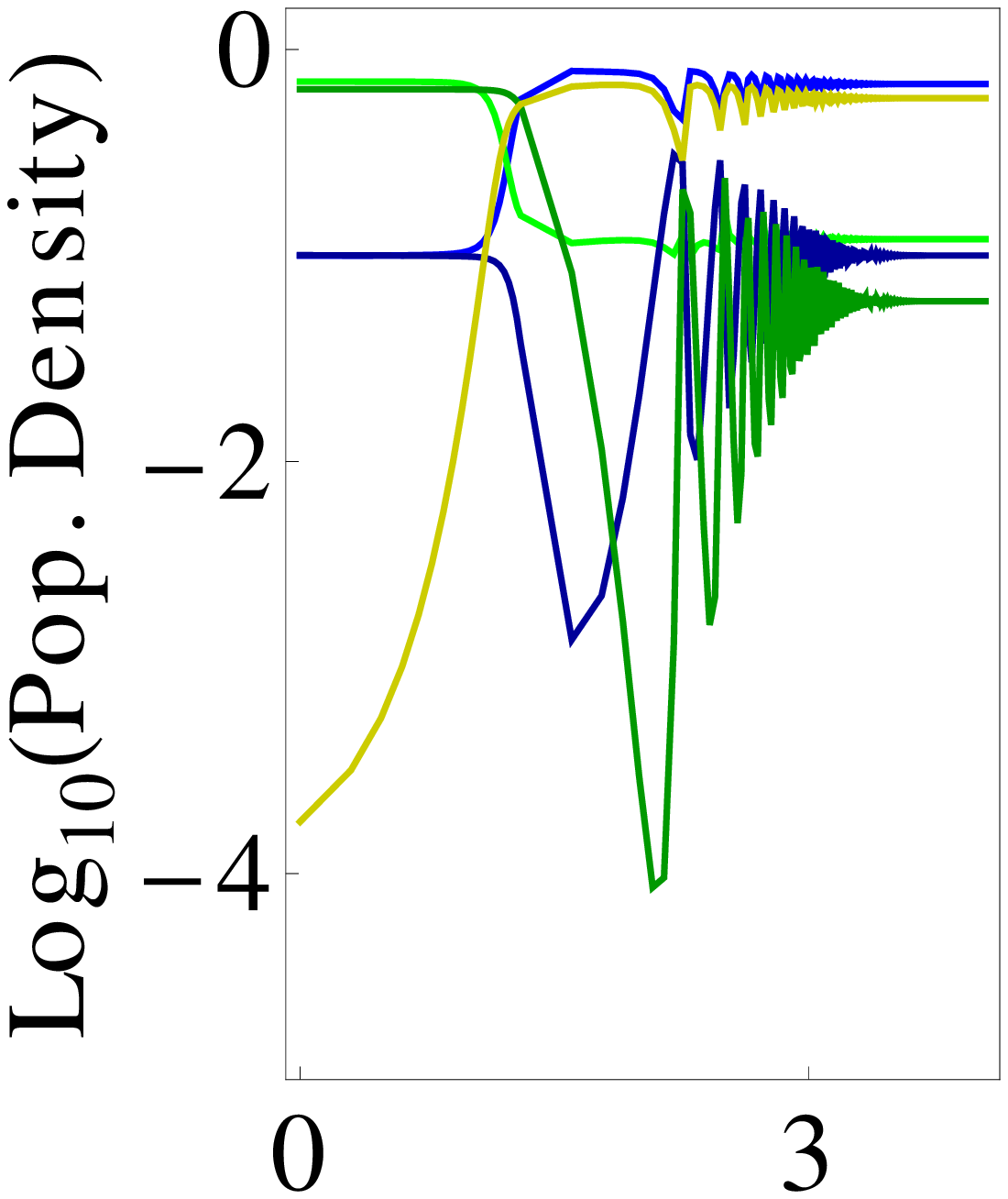}}
\put( 85,-65){\includegraphics[height=2.9cm,angle=0,trim= 3.2cm 0cm 0pt 0pt,clip]{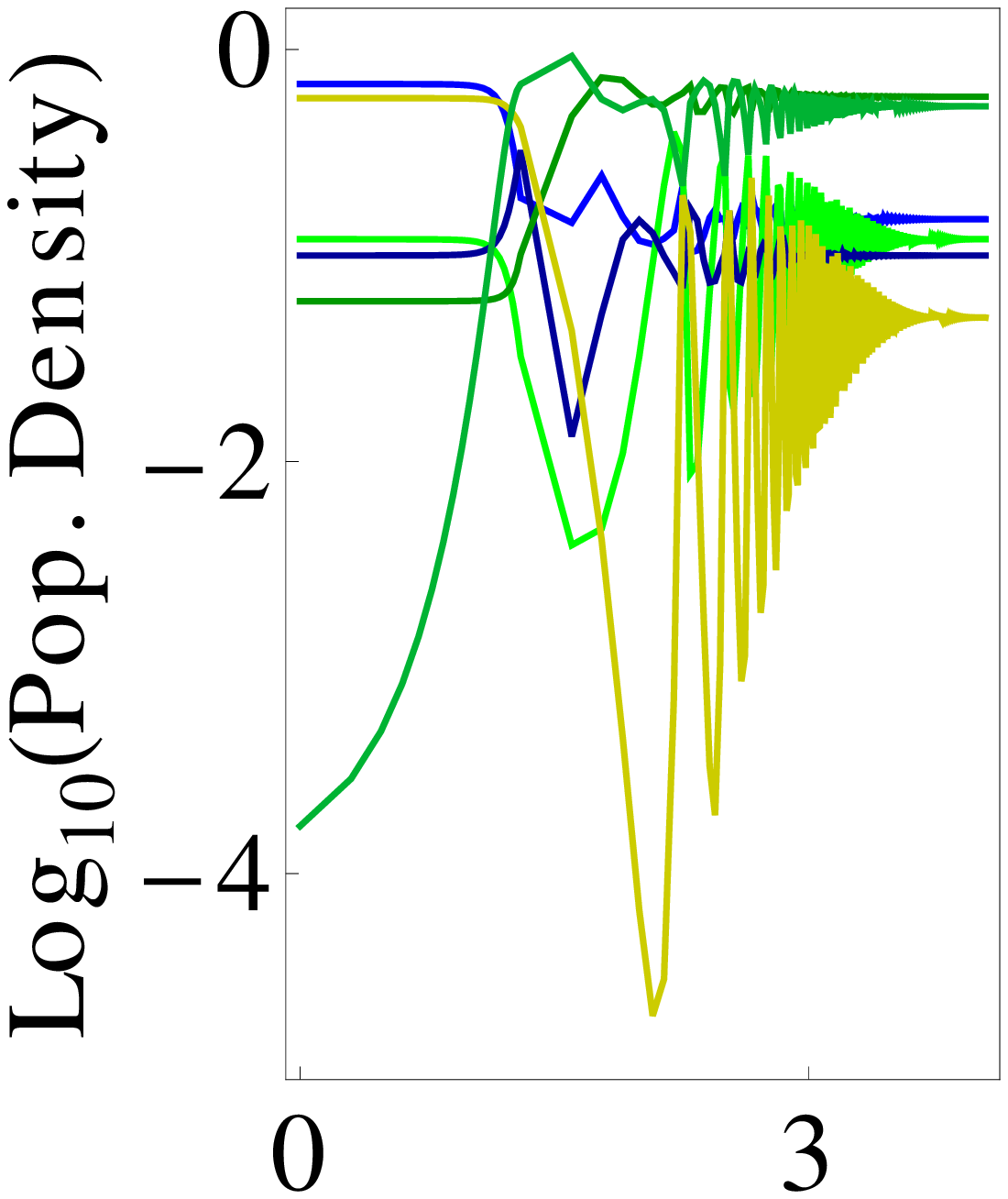}}
\put( 116.75,-65){\includegraphics[height=2.9cm,angle=0,trim= 3.2cm 0cm 0pt 0pt,clip]{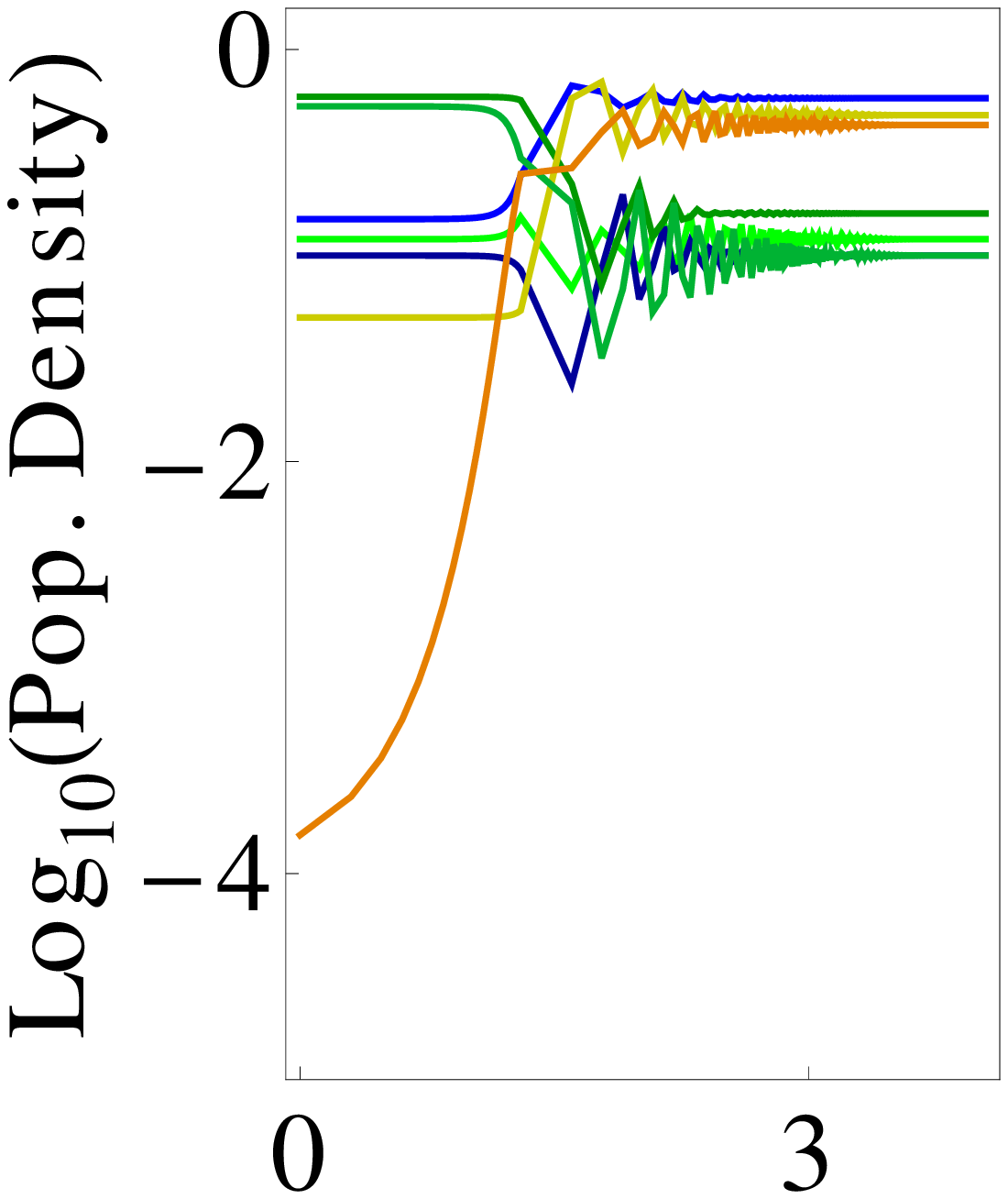}}
\put( 148,-65){\includegraphics[height=2.9cm,angle=0,trim= 3.2cm 0cm 0pt 0pt,clip]{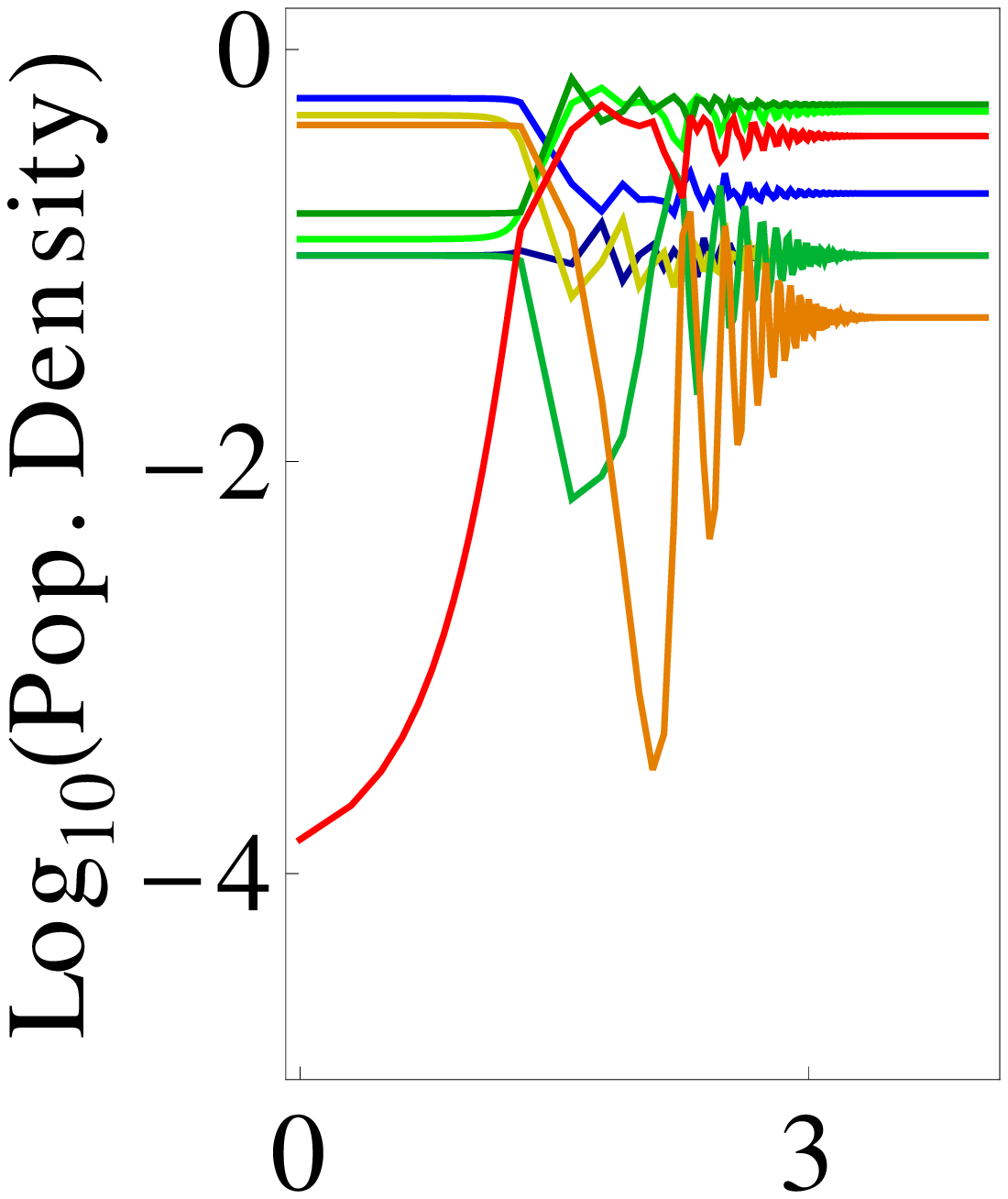}}
\put(35,-67){\small Log$_{10}$(Time)}
\end{overpic}
\vspace*{100pt}
\caption{\doublespacing{\bf Food web assembly.}  
{\bf a}, Consumer limited food web and its interaction matrix. The symbols ``$\times$'' mark nonzero entries and circles a path through the matrix. Basic nutrient is shown as a gray hexagon, whereas species on subsequent trophic levels are shown with coloured circles.
Small circles highlight limitations by consumers. Shaded ovals indicate possible pairing of species. 
{\bf b}, Possible assembly of the food web in (a) with labels for $N_o$+$N_e$ and $N_o$$-$$N_e$.
We set all growth and coupling constants equal to unity, but consider fine-tuned decay coefficients $\alpha \ll 1$ \cite{haerter2015inprep}.
Note the transitions between biomass limiting states.
Sizes of circles indicate approximate densities of species in the different states. Link shown in gray was set to zero in the numerical simulations in (c).
{\bf c}, Species added one-by-one as shown in (b) and granted small initial population density ($10^{-4}$). After each addition the system is integrated until steady state is reached. 
In the plot, colors of curves denote species of similar colors in the panels of (b). 
Note the double-logarithmic axis-scaling in (c). Time in each panel is relative to the time of introduction of the new species.
}
\label{fig:interaction_matrix}
\end{center}
\end{figure*}
\end{center}

\begin{center}
\begin{figure*}
\begin{center}
\begin{overpic}[width=5cm,angle=-90,trim= 0cm 0pt 0pt 0pt,clip]{dummy.eps}
\put(-40,10){\includegraphics[width=7.0cm,angle=0]{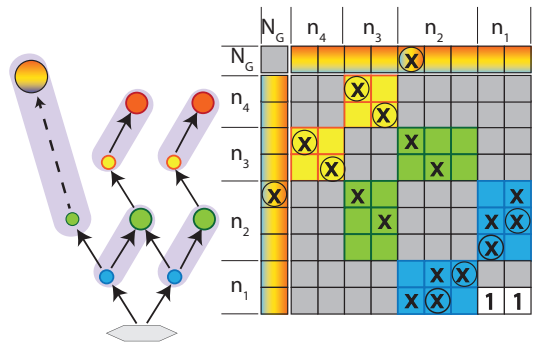}}
\end{overpic}
\vspace*{0pt}
\caption{\doublespacing{\bf Inclusion of omnivory.}  
Food web and corresponding interaction matrix for several species with sharp trophic levels and a single generalist omnivore. 
We only show its paired link in the network plot, but indicate all its interactions in the matrix.
}
\label{fig:interaction_matrix_omnivore}
\end{center}
\end{figure*}
\end{center}

\begin{center}
\begin{figure*}
\begin{center}
\begin{overpic}[width=5cm,angle=-90,trim= 0cm 0pt 0pt 0pt,clip]{dummy.eps}
\put(-80,-60){\includegraphics[width=13.4cm,angle=0]{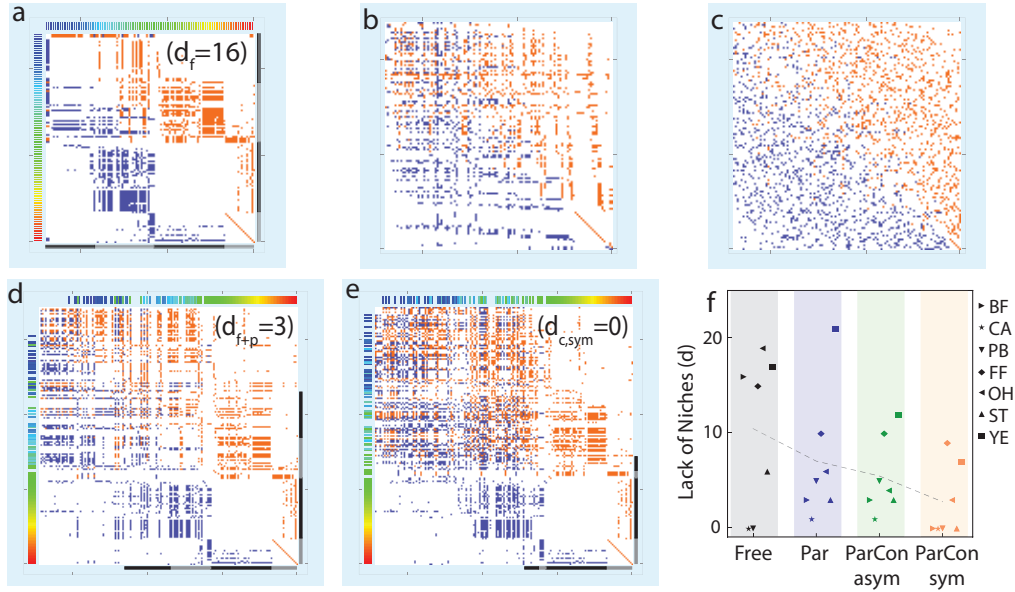}}
\end{overpic}
\vspace*{100pt}
\caption{\doublespacing{\bf Comparison to existing models and data analysis.}  
{\bf a}, Bahia Falsa free-living food web, with blue and red matrix elements for predator respectively prey dependency. Trophic levels indicated by gray and black bars, whereas color coding along the left and upper edge labels chain length of free-living species (increase from red to blue shades).
{\bf b}, Niche model simulation of the Bahia Falsa free-living food web.
{\bf c}, Cascade model simulation of the Bahia Falsa free-living food web.
{\bf d}, As (a) but with parasites (``Par'') and with colors for free-living species carried over from (a). Remaining species are parasites.
{\bf e}, As (d) but additionally including symmetric concomitant links (``ParCon sym'').
{\bf f}, Lack of niches ($d$) for seven empirical food webs \cite{dunne2013parasites}. Labels mark the sub-webs of free-living species (``Free``), including also parasite links (''Par``), asymmetric (''ParCon asym``) and several symmetric concomitant links (''ParCon sym``). Dashed line connects averages in these categories.
Rank deficiencies for free living ($d_f$), free-living and parasite species ($d_{f+p}$) as well as additional concomitant links ($d_{c,sym}$) marked in (a), (d), and (e), respectively. (Analysis details and abbreviations: see Methods).
}
\label{fig:emp_interaction_matrix}
\end{center}
\end{figure*}
\end{center}

\begin{center}
\begin{figure*}
\begin{center}
\begin{overpic}[width=5cm,angle=-90,trim= 0cm 0pt 0pt 0pt,clip]{dummy.eps}
\put(-40,-125){\includegraphics[width=7.4cm,angle=0]{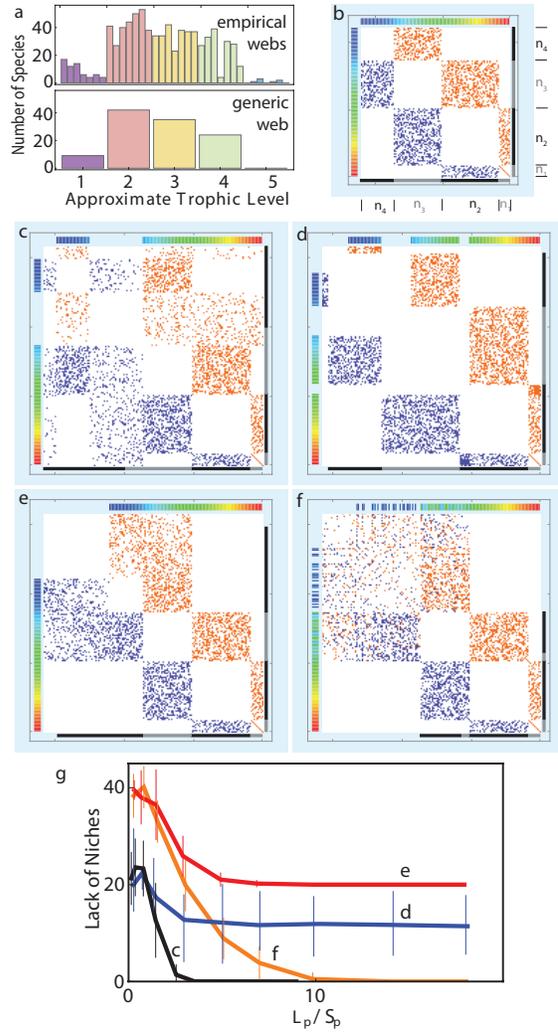}}
\end{overpic}
\vspace*{180pt}
\caption{\doublespacing{{\bf Simulations of different food web matrices.} 
{\bf a}, Barplots indicate distributions of node richness for each approximate trophic level in the seven empirical foodwebs and a generic foodweb derived by averaging the empirical node richnesses in each trophic level.
{\bf b}, Interaction matrix corresponding to the generic food web, containing 110 free-living and 47 parasite species (Details: Methods and Sec.~S10).
{\bf c}, Addition of parasites that form random links to any existing free-living species. 
{\bf d}, Addition of parasites that are confined to consumer at a specific trophic level. 
{\bf e}, Similar to (c) but with the restriction of parasites consuming only free-living species at levels 3 and 4. 
{\bf f}, Similar to (e) but with additional parasite-parasite interactions (hyperparasitism), approximately 5 percent of parasite links are from parasite to parasite. 
Note the color coding along the edges of the matrices in (c)---(f), chosen as in Fig.~\ref{fig:emp_interaction_matrix}a,d,e.
{\bf g}, The lack of niches, i.e. rank deficiency, as a function of the number of links per parasite for each of the four cases described in (c)---(f).
}}
\label{fig:sim_Matrices}
\end{center}
\end{figure*}
\end{center}

\end{document}